\documentclass[preprint,tightenlines,aps,prd,showpacs]{revtex4}
\usepackage{graphicx}
\usepackage{amsmath}
\usepackage{bm}

\begin{document}

\title{\mbox{}\\[10pt]
Low-energy Universality and the \\
New Charmonium Resonance at 3870 MeV}

\author{Eric Braaten and Masaoki Kusunoki}
\affiliation{
Physics Department, Ohio State University, Columbus, Ohio 43210}

\date{\today}
\begin{abstract}
The recently-discovered narrow charmonium resonance near 3870 MeV
is interpreted as a hadronic molecule 
whose constituents are the charm mesons $D^0$ and $\bar D^{*0}$
or $\bar D^0$ and $D^{*0}$.
Because of an accidental fine-tuning 
of the molecule to very near the 
$D^0 \bar D^{*0}$ threshold, it has some universal 
properties that are completely determined by the unnaturally 
large $D^0 \bar D^{*0}$ scattering length $a$.
Its narrow width can be explained by the suppression 
by a factor of $1/a$ of decay modes other than 
the decay of a constituent $\bar D^{*0}$ or $D^{*0}$.
Its production rates are also suppressed by a factor of $1/a$.  
A particularly predictive mechanism for generating the 
large scattering length is the accidental fine-tuning of 
a P-wave charmonium state to the 
$D^0 \bar D^{*0}$ threshold.  
\end{abstract}



\maketitle


\section{Introduction}

The discovery of the $J/\psi$ and other charmonium resonances 
in 1974 played a crucial role in the construction 
of the gauge theories of the strong and electroweak 
forces that constitute the Standard Model of particle physics.
Subsequent studies of the spectroscopy, decays, and production 
of both charmonium and bottomonium resonances have played important 
roles in the development of quantum chromodynamics (QCD), 
the gauge theory of the strong interactions.  
The recent unexpected discovery by the Belle collaboration 
of a narrow charmonium resonance near 3.87 GeV \cite{Choi:2003ue} 
has presented a new challenge to our understanding of QCD.

The new charmonium state $X(3870)$ was discovered in electron-positron 
collisions through the $B$-meson decay $B^\pm  \to K^\pm X$
followed by the decay $X  \to J/\psi \pi^+ \pi^-$.
Its mass was measured to be $M_X=3872.0 \pm 0.6 \pm 0.5$ MeV
\cite{Choi:2003ue}.
It is narrow compared to other charmonium states  
above the threshold for decay into two charm mesons:
the upper bound on the width is $\Gamma_X < 2.3$ MeV.
The discovery has been confirmed by the CDF collaboration 
who observed $X$ through $J/\psi \pi^+ \pi^-$ events in 
proton-antiproton collisions and measured its mass to be
$M_X = 3871.4 \pm 0.7 \pm 0.4$ MeV \cite{CDF}.

There have been several recent papers discussing the possible 
interpretations of the $X(3870)$ 
\cite{Tornqvist:2003na,Close:2003sg,Pakvasa:2003ea,Voloshin:2003,YMW:2003,Wong:2003}.
The most conventional interpretations are previously undiscovered 
states in the charmonium spectrum, such as 
one of the lowest D-wave states with spin/parity/charge-conjugation
quantum numbers $J^{PC}= 2^{--}$ or $2^{-+}$ 
or one of the first excited P-wave states 
with $J^{PC}= 1^{++}$ or $1^{+-}$.   A more exotic possibility is
a ``hybrid charmonium'' state in which a gluonic mode 
has been excited.  Another possibility, a $D^0 \bar D^{*0}$  
or $\bar D^0 D^{*0}$ molecule,
is motivated by the fact that the $X(3870)$ is
extremely close to the threshold $3871.2 \pm 0.7$ MeV for decay
into the charmed mesons $D^0 \bar D^{*0}$.  
The possibility of hadronic molecules formed from charm mesons was 
suggested long ago \cite{Voloshin:ap,DeRujula:1976qd}. 
The most favorable channels for forming a molecule 
from the pion-exchange interaction are the P-wave
channel with $J^{PC}= 0^{-+}$ and the S-wave channel with 
$J^{PC}= 1^{++}$ \cite{Tornqvist:2003na}.

In this paper, we explore the consequences of identifying $X(3870)$ 
as an S-wave $D^0 \bar D^{*0} /\bar D^0 D^{*0}$ molecule.  
The tiny binding energy of the molecule implies 
that the $D^0 \bar D^{*0}$ scattering length $a$ is unnaturally large.  
The molecule therefore has properties that depend on $a$ but are 
insensitive to other details of the interactions of $D^0$
and $\bar D^{*0}$, a phenomenon called ``low-energy universality."  
In Section~\ref{sec:LEU}, we discuss the implications of 
low-energy universality for the wavefunction of $X$ 
and describe two possible mechanisms for
generating the large $D^0 \bar D^{*0}$ scattering length.  
In Section~\ref{sec:Decays}, we discuss the implications of 
low-energy universality for decays of $X$.  
One mechanism for generating the large $D^0 \bar D^{*0}$ 
scattering length, the fine-tuning of the energy of 
a P-wave charmonium state
to the $D^0 \bar D^{*0}$ threshold, 
gives a particularly distinctive pattern of branching fractions.  
Low-energy universality gives highly nontrivial predictions 
for 3-body systems, such as $D^0 D^0 \bar D^{*0}$.  
Unfortunately, as shown in Section~\ref{sec:Efimov}, 
the spectacular possibility of shallow $D^0 D^0 \bar D^{*0}$ molecules 
called Efimov states can be excluded.  In Section~\ref{sec:Model}, 
we present a nonrelativistic effective field theory 
that illustrates the two mechanisms for generating a large 
$D^0 \bar D^{*0}$ scattering length.  Our results are summarized in
Section~\ref{sec:Summary}.

\section{Low-energy Universality}
\label{sec:LEU}

We will assume that the closeness of $M_X$ to the $D^0 \bar D^{*0}$
threshold is no accident and that
$X(3870)$ is indeed a  hadronic molecule 
whose constituents are the charm mesons $D^0$ and $\bar D^{*0}$
or $\bar D^0$ and $D^{*0}$.  
What makes this molecule unique among all the hadrons that can be
interpreted as 2-body bound states of other 
hadrons is its extremely small binding energy.
If the low-energy interaction between two hadrons is mediated by 
pion exchange, the natural scale for the binding energy of a molecule 
composed of the two hadrons is $m_\pi^2/(2 m_{\rm red})$,
where $m_{\rm red}$ is their reduced mass.  
The natural energy scale for a  $D^0 \bar D^{*0}$ molecule 
is about 10 MeV.
The binding energy of the $X(3870)$ (which is positive by definition)
has been measured to be $B_2 = -0.5 \pm 0.9$ MeV.  
Thus it is likely to be less than 0.4 MeV,
which is much smaller than the natural energy scale.
The only other two-body bound state of hadrons whose binding energy
is known to be small compared to the natural energy scale is the deuteron.
Its binding energy is 2.4 MeV, which is small compared to the 
natural energy scale of 20 MeV for a $pn$ molecule.

We will further assume that $X(3870)$ is an S-wave bound state of
$D^0 \bar D^{*0}$ or $\bar D^0 D^{*0}$,
because this has particularly interesting implications. 
Since the constituents have $J^P$ quantum numbers $0^-$ and $1^-$,
the $J^{PC}$ quantum numbers of the molecule 
must be $1^{++}$ or $1^{+-}$.
The interaction between $D^0$ and $\bar D^{*0}$ 
at energies less than $m_\pi^2/m_D \approx 10$ MeV is dominated by
the S-wave channel and can be described 
by a single parameter: the S-wave $D^0 \bar D^{*0}$ 
scattering length $a$.  
A shallow S-wave bound state implies an S-wave scattering length 
that is large compared to the natural length scale $1/m_\pi$.
The low-energy few-body observables for nonrelativistic particles 
with short-range interactions and a large scattering length 
have universal features that are insensitive to the details of the 
mechanism that generates the large scattering length.  
This phenomenon is called {\it low-energy universality}.
If $a>0$, the simplest universal prediction is that there is a shallow 
2-body bound state whose binding energy $B_2$ for sufficiently
large $a$ approaches
\begin{eqnarray}
B_2 \longrightarrow  {1 \over 2 m_{\rm red} a^2},
\label{B2}
\end{eqnarray}
where $m_{\rm red}$ is the reduced mass of the two constituents. 
If the binding energy of the $X(3870)$ were measured,
the $D^0 \bar D^{*0}$ scattering length $a$ could be predicted 
using (\ref{B2}) with 
$m_{\rm red} = m_{D^0} m_{D^{*0}}/(m_{D^0}+m_{D^{*0}})$. 
For example, if the binding energy of $X(3870)$ were 0.5 MeV or 0.1 MeV, 
the scattering length would be 6.3 fm or 14.2 fm, respectively.
These are both much larger 
than the natural length scale $1/m_\pi = 1.5$ fm.

Low-energy universality has other implications for the
interpretation of $X(3870)$ as  
a $D^0 \bar D^{*0}$/$\bar D^0 D^{*0}$ molecule.
There is a universal prediction for the $D^0 \bar D^{*0}$ 
or $\bar D^0 D^{*0}$ wavefunction:
\begin{eqnarray}
\psi(r) \longrightarrow (2 \pi a)^{-1/2} {\exp(-r/a) \over r}.
\label{psi}
\end{eqnarray}
Voloshin has exploited this universal wavefunction to
calculate the momentum distributions for the decays 
$X \to D^0 \bar D^0 \pi^0$ and $X \to D^0 \bar D^0 \gamma$ 
\cite{Voloshin:2003}.
There are also components of the wavefunction that correspond
to other hadronic states with the same $J^{PC}$ quantum numbers.
If $J^{PC}=1^{++}$, they include
the P-wave charmonium states $\chi_{c1}(1P)$ and $\chi_{c1}(2P)$.
If $J^{PC}=1^{+-}$, they include
the P-wave charmonium states $h_c(1P)$ and $h_c(2P)$.
In either case, they also include $D^+ D^{*-}$/$D^- D^{*+}$ states 
and $D^{*0} \bar D^{*0}$ states.
In an appropriate hadronic basis, the quantum state for $X(3870)$
can be written
\begin{eqnarray}
| X \rangle &=& Z_{DD^*}^{1/2} \int {d^3p \over (2 \pi)^3} \, \tilde \psi(p)
{1 \over \sqrt{2}}
\left(
| D^0({\bf p}) \bar D^{*0}(-{\bf p}) \rangle
\pm | \bar D^0({\bf p}) D^{*0}(-{\bf p}) \rangle \right),
\nonumber
\\
&& + \sum_H Z_H^{1/2} | H \rangle.
\label{wavefcn}
\end{eqnarray}
where $\tilde \psi(p)$ is the Fourier transform of the 
$D^0 \bar D^{*0}$/$\bar D^0 D^{*0}$ wavefunction
and the sign $\pm$ is determined by the charge conjugation 
quantum number $C=\pm 1$.
The other hadronic states $H$ can be discrete states, 
such as $\chi_{c1}(2P)$ or $h_c(2P)$, or continuum states, 
such as $D^+({\bf p}) D^{*-}(-{\bf p})$.
The probability factors 
$Z_{DD^*}$ and $Z_H$ are real and positive,
and they add up to 1. 
Low-energy universality implies that as the scattering length $a$
increases, the probabilities for states other than
$D^0 \bar D^{*0}$ or $\bar D^0 D^{*0}$ decrease as $1/a$
and the $D^0 \bar D^{*0}$/$\bar D^0 D^{*0}$ wavefunction approaches
(\ref{psi}).
In the limit $a \to \infty$,
the state becomes a pure $D^0 \bar D^{*0}$/$\bar D^0 D^{*0}$
molecule.

A scattering length that is large compared to the natural length 
scale necessarily requires a fine-tuning.
In the case of the $D^0 \bar D^{*0}$ molecule, the fine-tuning
parameters can be identified with the up and down quark masses 
$m_u$ and $m_d$.  The masses of $D^0$ and $\bar D^{*0}$ are sensitive 
to $m_u$, because these hadrons contain an up quark as a constituent.
The $D^0 \bar D^{*0}$ potential is sensitive to $m_u$ and $m_d$
through the pion mass.
There are two distinct mechanisms for generating a large 
$D^0 \bar D^{*0}$ scattering length.
The first mechanism is a fine-tuning of 
parameters that have a large effect on the 
$D^0 \bar D^{*0}$/$\bar D^0 D^{*0}$ channel
without significantly affecting other channels.
This could be a fine-tuning of the range and depth of the 
$D^0 \bar D^{*0}$ potential so that there is a
bound state very close to threshold and thus a large scattering 
length.  Equivalently, it could be a fine-tuning of the masses of the $D^0$
and $\bar D^{*0}$ to obtain a bound state very close to threshold 
in the $D^0 \bar D^{*0}$ potential.
This mechanism requires the quantum numbers of $X$ to be 
$J^{PC} = 1^{++}$, because this is the only S-wave channel 
for which the potential due to pion exchange 
is sufficiently attractive to produce a bound state \cite{Tornqvist:2003na}.
In the limit $a \to \infty$, the probabilities for components 
of the wavefunction 
other than $D^0 \bar D^{*0}$ or $\bar D^0 D^{*0}$
scale as $1/a$.
This will be illustrated in Section~\ref{sec:Model} 
using an explicit field theory model.
If the energy gap $\nu_H$ between the state $H$ in 
(\ref{wavefcn}) and the $D^0 \bar D^{*0}$ threshold 
is much greater than the natural energy scale $m_\pi^2/m_{\rm red}$,
a more complete estimate of the dimensionless suppression 
factor in $Z_H$ is $m_\pi^3/(m_{\rm red}^2 \nu_H^2 a)$. 
For most channels, the energy gap $\nu_H$
is much larger than the natural energy scale.  
For example, the energy gap for $\chi_{c1}(1P)$
is $\nu_H = - 360$ MeV.  If $a m_\pi \gg 1$, the suppression factor 
$m_\pi^3/(m_{\rm red}^2 \nu_H^2 a)$
is 1/1600 for $B_2 = 0.5$ MeV and 1/3500 for $B_2 = 0.1$ MeV.
If this is the correct mechanism for generating the large scattering length,
we can probably neglect all components of the 
wavefunction other than $D^0 \bar D^{*0}$/$\bar D^0 D^{*0}$ 
and set $Z_{DD^*} \approx 1$.

A second mechanism for a large $D^0 \bar D^{*0}$ scattering length 
is an accidental fine-tuning of one of the P-wave 
charmonium states $\chi_{c1}(2P)$ or $h_c(2P)$
to the $D^0 \bar D^{*0}$ threshold.
$X(3870)$ will have the same $J^{PC}$ quantum numbers 
as the charmonium state: $1^{++}$ in the case of $\chi_{c1}(2P)$
and $1^{+-}$ in the case of $h_c(2P)$.
This mechanism is analogous to the Feshbach resonances
\cite{TVS93} that can be used to control the scattering lengths 
for atoms by adjusting the 
magnetic field \cite{Inouye98,Courteille98,Roberts98}.
Feshbach resonances are currently being used to tune the scattering lengths 
for atoms to arbitrarily large values in order to study 
Bose-Einstein condensates of bosonic atoms
and degenerate gases of fermionic atoms
in the strongly-interacting regime.
In the case of the $D^0 \bar D^{*0}$ molecule, 
the fine-tuning parameter can be identified as
$m_u$, which can shift the $D^0$ and $\bar D^{*0}$ masses, 
thus changing the energy gap $\nu$ between 
the $\chi_{c1}(2P)$ or $h_c(2P)$ and the $D^0 \bar D^{*0}$ threshold.
In potential models, which ignore the coupling of charmonium states 
to continuum channels such as $D \bar D$ and $D \bar D^*$,
the estimates of the energy gap for $\chi_{c1}(2P)$ or $h_c(2P)$
are both $\nu \approx 90$ MeV
\cite{Buchmuller:1980su,Godfrey:xj}.
The predictions of these models for the mass of the 1D state
$\psi(3770)$ with $J^{PC} = 1^{--}$ are too large by about 50 MeV,
so they are also likely to overpredict the 
masses of the $\chi_{c1}(2P)$ and $h_c(2P)$.
The error presumably arises mostly from the neglect 
of coupled-channel effects, which are sensitive to the light quark masses.  
If the coupled-channel effects shift the 2P charmonium states
down by about 90 MeV relative to the $D \bar D^*$ threshold,
they could fortuitously tune the energy gap 
$\nu$ for $\chi_{c1}(2P)$ or $h_c(2P)$
to be smaller than the natural 
low-energy scale $m_\pi^2/m_D \approx 10$ MeV associated with pion 
exchange.  In this case, a resonant interaction between 
the $\chi_{c1}(2P)$ or $h_c(2P)$ and $D^0 \bar D^{*0}$ states generates a large 
$D^0 \bar D^{*0}$ scattering length $a$ that increases as $1/\nu$. 
If  $a>0$, there is a shallow bound state whose
binding energy approaches (\ref{B2}) as $\nu \to 0$. 
In the expression (\ref{wavefcn}) for the quantum state,
the hadrons should be interpreted as those in the absence of the 
fine-tuning that generates the resonant interaction. 
In the limit $a \to \infty$, the probability $Z_\chi$  
for $\chi_{c1}(2P)$ or $Z_h$ for $h_c(2P)$
scales as $1/a$, as do the probabilities $Z_H$
for all other channels besides $D^0 \bar D^{*0}$/$\bar D^0 D^{*0}$.
This will be illustrated in Section~\ref{sec:Model}
using an explicit field theory model.
The probability $Z_\chi$ or $Z_h$ includes a 
dimensionless suppression factor $1/(a m_\pi)$, whose value
is about 1/4.3 for $B_2 = 0.5$ MeV and  1/9.7 for $B_2 = 0.1$ MeV.
There may also be further suppression from a small numerical coefficient
associated with Zweig's rule, because the processes 
$\chi_{c1}(2P) \to D^0 \bar D^{*0}$ and $h_c(2P) \to D^0 \bar D^{*0}$
require the creation of a light quark-antiquark pair.
If this is the correct mechanism for generating the large scattering length,
we can probably neglect all components of the 
wavefunction other than $D^0 \bar D^{*0}$/$\bar D^0 D^{*0}$ 
and $\chi_{c1}(2P)$ or $h_c(2P)$.
We can then set $Z_{DD^*} \approx 1-Z_\chi$ in the case $J^{PC} = 1^{++}$
and $Z_{DD^*} \approx 1-Z_h$ in the case $J^{PC} = 1^{+-}$.

\section{Decays}
\label{sec:Decays}

An important requirement for any interpretation of the 
$X(3870)$ is that it provide an explanation for its narrow width.
The upper bound $\Gamma_X < 2.3$ MeV implies that the width of 
$X$ is more than an order of magnitude smaller than 
that of the D-wave state $\psi(3770)$.
According to our interpretation, $X$ is below the $D^0 \bar D^{*0}$ 
threshold and it therefore cannot decay into $D^0 \bar D^{*0}$.  
Its quantum numbers 
$J^{P} = 1^{+}$ forbid a decay into $D^0 \bar D^0$ or $D^+ D^-$.
It can however decay into $D^0 \bar D^0 \pi^0$
or $D^0 \bar D^0 \gamma$ by the decay of a constituent 
$D^{*0}$ or $\bar D^{*0}$ of the molecule.
It can also decay
into a lighter charmonium state by a radiative or hadronic transition 
or into light hadrons via a process in which the charm quark 
and antiquark annihilate.  

We first consider the contribution to the width $\Gamma_X$ from the 
decay of a constituent $D^{*0}$ into $D^0 \pi^0$ or $D^0 \gamma$.
The width $\Gamma[D^{*0} \to D^0 \pi^0]$ can be deduced from the 
measured width of the $D^{*+}$, the branching fraction for
$D^{*+} \to D^0 \pi^+$ or $D^{*+} \to D^+ \pi^0$, and isospin symmetry:
$\Gamma[D^{*0} \to D^0 \pi^0] = 31 \pm 7$ keV.
We have treated the difference between the branching fraction for
$D^{*+} \to D^0 \pi^+$ and twice the branching fraction for
$D^{*+} \to D^+ \pi^0$ as a systematic error associated with 
isospin symmetry breaking.
The width $\Gamma[D^{*0} \to D^0 \gamma]$ can then be deduced from 
the measured branching fractions for $D^{*0} \to D^0 \pi^0$
and $D^{*0} \to D^0 \gamma$:
$\Gamma[D^{*0} \to D^0 \gamma] = 19 \pm 5$ keV.
The contributions to the width $\Gamma_X$ from these decays are
\begin{subequations}
\begin{eqnarray}
\Gamma[X \to D^0 \bar D^0 \pi^0] &=& 
Z_{D D^*} C_\pi \Gamma[D^{*0} \to D^0 \pi^0],
\label{Gam-pi}
\\
\Gamma[X \to D^0 \bar D^0 \gamma] &=& 
Z_{D D^*} C_\gamma \Gamma[D^{*0} \to D^0 \gamma].
\label{Gam-gam}
\end{eqnarray}
\end{subequations}
The factors $C_\pi$ and $C_\gamma$ take into account 
interference between the decay of
$\bar D^{*0}$ from the $D^0 \bar D^{*0}$
component of the wavefunction and the decay of
$D^{*0}$ from the $\bar D^0 D^{*0}$ component. 
If the charge conjugation quantum number of $X$ is $C = +1$,
there is constructive interference in the decay 
$X \to D^0 \bar D^0 \pi^0$ and destructive interference 
in the decay $X \to D^0 \bar D^0 \gamma$.
If $C = -1$,  the pattern is reversed \cite{Voloshin:2003}.
If $C = +1$ (or $-1$), the coefficient $C_\pi$ ranges from
about 1.5 (or 0.13) if the binding energy $B_2$ is 0.5 MeV
to about 2.2 (or 0.56) if $B_2=0.1$ MeV and to 2 if $B_2=0$.
The coefficient $C_\gamma$ ranges from
about 0.58 (or 3.42) if $B_2 = 0.5$ MeV 
to about 1.36 (or 2.64) if $B_2 = 0.1$ MeV
and to 2 if $B_2=0$ \cite{Voloshin:2003}. 
The probability factor $Z_{D D^*}$ is close to 1.
Thus the lower bound on the width 
provided by the sum of (\ref{Gam-pi}) and (\ref{Gam-gam})
ranges from 
$56 \pm 11$ keV (or $69 \pm 16$  keV) if $B_2 = 0.5$ MeV 
to $92 \pm 17$  keV (or $67 \pm 13$  keV)
if $B_2 = 0.1$ MeV to $99 \pm 17$ keV if $B_2=0$. 

There can also be decays that proceed through the decay of 
a constituent $D^{*+}$ or $D^{*-}$ 
from the $D^- D^{*+}$ or $D^+ D^{*-}$
component of the wavefunction.  The decays of $X$ into 
$D^+ D^- \pi^0$, $D^0 D^- \pi^+$, or $D^+ \bar D^0 \pi^-$
are forbidden by phase space.  The decay $X \to D^+ D^- \gamma$
is allowed, but it is suppressed by the small decay rate 
$\Gamma[D^{*+} \to D^+ \gamma] = 1.5 \pm 0.5$ keV
and also by the small probability for the $D^- D^{*+}$/$D^+ D^{*-}$
component of the wavefunction.
Thus the contribution to the width $\Gamma_X$ from the 
$D^- D^{*+}$ /$D^+ D^{*-}$
component of the wavefunction can be neglected.

The remaining decay channels of the $X(3870)$
are radiative transitions to lower charmonium states 
such as $X \to \psi(2S) \gamma$ or $X \to \eta_c(2S) \gamma$, 
hadronic transitions to lower charmonium states such as the 
discovery mode $X \to J/\psi \pi^+ \pi^-$, 
and annihilation decays into light hadrons such as $X \to p \bar p$.
These decays proceed through the 
short-distance part of the $D^0 \bar D^{*0}$/$\bar D^0 D^{*0}$ 
wavefunction or through other components 
of the wavefunction, such as $\chi_{c1}(2P)$ or $h_c(2P)$.
All these contributions to the decay rate 
are suppressed by a factor of $1/a$ and
go to 0 as the binding energy of $X$ goes to 0.

If the large $D^0 \bar D^{*0}$ scattering length 
arises from an accidental fine-tuning within the 
$D^0 \bar D^{*0}$/$\bar D^0 D^{*0}$ channel with $J^{PC} = 1^{++}$,
the probability $Z_H$ for 
a component of the wavefunction 
with a large energy gap $|\nu_H| \gg m_\pi^2/m_{\rm red}$
is suppressed by $m_\pi^3/(m_{\rm red}^2 \nu_H^2 a)$.
Thus the radiative transitions, hadronic transitions, 
and annihilation decays of $X(3870)$ are dominated by the 
$D^0 \bar D^{*0}$/$\bar D^0 D^{*0}$ component of the wavefunction.
For example, the amplitudes for radiative and hadronic transitions
to $J/\psi$  can be approximated by
\begin{subequations}
\begin{eqnarray}
{\cal A}[X \to J/\psi + \gamma] &\approx& 
Z_{DD^*}^{1/2} \int {d^3p \over (2 \pi)^3} \, \tilde \psi(p)
\sqrt{2} {\cal A}[ D^0({\bf p}) \bar D^{*0}(-{\bf p}) \to J/\psi + \gamma],
\label{amp-rad}
\\
{\cal A}[X \to J/\psi + h] &\approx& 
Z_{DD^*}^{1/2} \int {d^3p \over (2 \pi)^3} \, \tilde \psi(p)
\sqrt{2} {\cal A}[ D^0({\bf p}) \bar D^{*0}(-{\bf p}) \to J/\psi + h],
\label{amp-had}
\end{eqnarray}
\label{decay-1}
\end{subequations}
where $h$ consists of light hadrons.
An example of a light hadronic state $h$ is the discovery channel 
$\pi^+ \pi^-$.  The light hadronic state $h$ must be odd under
charge conjugation.  Since $D^0$ and $\bar D^{*0}$ have isospin 
$1\over2$ and $J/\psi$ has isospin 0, $h$ can have isospin 0 or 1. 
For $p \ll m_\pi$, the momentum space wavefunction $\tilde \psi(p)$ 
can be approximated by its universality limit:
\begin{eqnarray}
\tilde \psi(p) \longrightarrow (2 \pi a)^{-1/2} {1 \over p^2 + 1/a^2}.
\label{psi-mom}
\end{eqnarray}
The momentum integrals in (\ref{decay-1}) 
are cut off at large momentum by the $p$-dependence of the 
transition amplitude for $D^0 \bar D^{*0} \to J/\psi + h$.
For this transition to occur, the heavy $c$ and $\bar c$ in the 
$J/\psi$ must be in the same momentum states as the $c$ and $\bar c$
in the $D^0$ and $\bar D^{*0}$.  Thus the transition amplitude 
includes a factor of the momentum space wavefunction 
$\tilde \psi_{J/\psi}(p)$ for the $J/\psi$.  
The momentum scale associated with this wavefunction
is $m_c v \approx 700$ MeV, where $v$ is the typical velocity 
of the charm quark in charmonium.
The amplitude on the 
right side of (\ref{decay-1}) therefore includes the overlap factor
$\int d^3p \, \tilde \psi(p) \, \tilde \psi_{J/\psi}(p)$.
The explicit factor of $a^{-1/2}$ in the 
universal wavefunction in (\ref{psi})
combines with a factor of $(m_c v)^{-1/2}$  from the integral
to give a dimensionless suppression factor.
Thus the decay rate scales as $1/(a m_c v)$ as $a \to \infty$.
If a phenomenological framework for calculating 
radiative and hadronic transitions of $D^0 \bar D^{*0}$ to quarkonium
were available, the rates for radiative and hadronic transitions of $X$
could be calculated using equations analogous to (\ref{decay-1}).

If the large $D^0 \bar D^{*0}$ scattering length 
arises from an accidental fine-tuning of the P-wave 
charmonium state $\chi_{c1}(2P)$ to the $D^0 \bar D^{*0}$ threshold, 
the radiative and hadronic transitions 
and the annihilation decays of $X(3870)$ can also proceed through
the $\chi_{c1}(2P)$ component of the wavefunction. 
These contributions to the decay rate are suppressed by the 
probability factor $Z_\chi$, which scales as $1/(a m_\pi)$. 
Although this suppression factor has the same power of $a$
as in the $D^0 \bar D^{*0}$ contribution,
the $\chi_{c1}(2P)$ contribution may be numerically larger 
because of a factor of $m_c v/m_\pi$ in the ratio of the suppression factors.
Thus decays of $X$ into modes that are possible final states of 
the decay of $\chi_{c1}(2P)$ are likely to be dominated 
by the $\chi_{c1}(2P)$ component of the wavefunction.
The rate for these decays will be given by the rate 
for the corresponding decays of the $\chi_{c1}(2P)$ 
in the absence of the fine-tuning multiplied by the 
probability factor $Z_\chi$.  For example, 
the decay rates for radiative transitions to $J/\psi$ 
and for hadronic transitions to $J/\psi$ via the emission of
a light hadronic state with total isospin 0 are
\begin{subequations}
\begin{eqnarray}
\Gamma[X \to J/\psi + \gamma] &\approx& 
Z_\chi \Gamma[\chi_{c1}(2P) \to J/\psi + \gamma],
\label{Gam-rad}
\\
\Gamma[X \to J/\psi + h_{I=0}] &\approx& 
Z_\chi \Gamma[\chi_{c1}(2P) \to J/\psi + h_{I=0}].
\label{Gam-had}
\end{eqnarray}
\label{decay-2}
\end{subequations}
The last factors in (\ref{Gam-rad}) and (\ref{Gam-had}) 
are the decay rates for $\chi_{c1}(2P)$ assuming that it has mass $m_X$ 
but ignoring the resonant interaction with $D^0 \bar D^{*0}$.
Hadronic transitions in which the light hadronic state $h$ has 
total isospin 1, such as $X \to J/\psi + \rho^0$, 
cannot proceed through the $\chi_{c1}(2P)$ component 
of the wavefunction. They must therefore be dominated by the 
$D^0 \bar D^{*0}$/$\bar D^0 D^{*0}$ component.
The amplitude for such a  transition can 
be approximated by (\ref{amp-had}).
There are well-developed phenomenological frameworks 
for calculating the radiative and hadronic transition rates 
\cite{Kuang:1981se,Godfrey:xj} 
for charmonium states such as $\chi_{c1}(2P)$.
If the rates for several radiative transitions or
hadronic transitions with isospin 0 were measured and found to be 
all suppressed relative to the predictions for $\chi_{c1}(2P)$
decays by a common factor $Z_\chi$, it would be strong evidence 
in favor of this fine-tuning mechanism.
The total width of $\chi_{c1}(2P)$ in the absence of
the resonant interaction with $D^0 \bar D^{*0}$
must be significantly larger than the width of 
$\chi_{c1}(1P)$, which is about 1 MeV,
since $\chi_{c1}(2P)$ has more decay channels 
and the decays have larger phase space.
However the suppression from the probability factor
$Z_\chi$ could reduce this contribution to the width
$\Gamma_X$ so that it is comparable to or even smaller than the 
contribution from the decay of $D^{*0}$ or $\bar D^{*0}$.

If the large $D^0 \bar D^{*0}$ scattering length 
arises from an accidental fine-tuning of the P-wave 
charmonium state $h_c(2P)$ to the $D^0 \bar D^{*0}$ threshold, 
the radiative and hadronic transitions 
and the annihilation decays of $X$ can also proceed through
the $h_c(2P)$ component of the wavefunction. 
The decay rates for radiative transitions 
and for hadronic transitions with total isospin 0
would be given by the corresponding decay rates of $h_c(2P)$
multiplied by a probability factor $Z_h$.

The identification of $X(3870)$ as a shallow S-wave molecule
also has implications 
for its production rate in high-energy collisions.
As the $D^0 \bar D^{*0}$ scattering length $a$ increases, 
the production rate decreases as $1/a$.
If the large value of $a$ 
arises from an accidental fine-tuning within the 
$D^0 \bar D^{*0}$ channel, the production will proceed 
primarily through the creation of $D^0$ and $\bar D^{*0}$
(or $\bar D^0$ and $D^{*0}$) with small relative momentum
of order $1/a$.  In this case, the suppression factor of $1/a$ 
in the production rate comes from the
factor $a^{-1/2}$ in the universal wavefunction (\ref{psi}).
If the large value of $a$ 
arises from an accidental fine-tuning of the P-wave 
charmonium state $\chi_{c1}(2P)$ to the $D^0 \bar D^{*0}$ threshold, 
the production rate may be dominated by production of the
$\chi_{c1}(2P)$. In this case, the suppression factor of $1/a$ 
comes from the probability $Z_\chi$ for the 
$\chi_{c1}(2P)$ component of $X$. 
Similarly, if the large value of $a$ 
arises from an accidental fine-tuning of $h_c(2P)$ 
to the $D^0 \bar D^{*0}$ threshold, 
the production rate is suppressed by a factor of $1/a$ 
from the probability $Z_h$ for the 
$h_c(2P)$ component of $X$.

\section{Absence of Efimov states}
\label{sec:Efimov}

The most remarkable predictions of low-energy universality, 
which were discovered by Efimov \cite{Efimov70},
occur in the 3-body sector.
At sufficiently low energies, the effective interaction between three
nonrelativistic particles with short-range forces
can be described by an effective potential $V_{\rm eff}(R)$
that depends only on the {\it hyperspherical radius} $R$,
which is a weighted average of the separations 
of the three particles \cite{NFJG01}.
If the scattering length is large compared to the range $\ell$ of the
force, the effective potential in the region 
$\ell \ll R \ll |a|$ is scale-invariant.  In the case of 
identical particles of mass $m$, 
the hyperspherical radius is just the root-mean-square
separation of the three pairs of particles and the scale-invariant
potential is
\begin{eqnarray}
V_{\rm eff}(R) \approx - {4 - \lambda_0 \over 2 m R^2},
\label{Veff}
\end{eqnarray}
where $\lambda_0$ is the minimum of the  nontrivial solutions to
\begin{eqnarray}
\sqrt{3} \lambda^{1/2} \cos(\pi \lambda^{1/2}/2) 
= 8 \sin(\pi \lambda^{1/2}/6).
\label{s0}
\end{eqnarray}
The minimum solution is $\lambda_0 = - s_0^2$, where 
$s_0 \approx 1.00624$.
In the {\it resonant limit} $a \to \infty$ in which the scattering length 
is tuned to be infinitely large,
the 2-body bound state has zero binding energy and
there are infinitely many arbitrarily-shallow 3-body bound states
called {\it Efimov states}.  If the particles are identical bosons,
the ratio of the binding energies of adjacent states approaches 
a universal constant $e^{2\pi/s_0} \approx 515.03$. 
The 3-body spectrum in the resonant limit 
has an asymptotic discrete scaling symmetry with discrete scaling factor
$e^{\pi/s_0} \approx 22.7$.  This symmetry is related to an infrared 
{\it renormalization group limit cycle} \cite{Braaten:2003eu}.
A limit that is more relevant to a physical problem with a 
large but finite scattering length is the {\it scaling limit} defined by 
$\Lambda \to \infty$ with $a$ fixed,
where $\Lambda$ is the natural momentum scale set by the inverse of the
range of the interaction.
In the scaling limit, the binding energies $B_3$ and $B_3'$ 
of the shallowest and next-to-shallowest Efimov states 
for identical bosons are in the intervals
$B_2 < B_3 < 6.75 B_2$ and $6.75 B_2 < B_3' < 1406 B_2$,
where $B_2 = 1/(ma^2)$ is the 2-body binding energy 
\cite{Braaten:2002jv}. 
If these binding energies are smaller than the natural energy scale
$\Lambda^2/m$, these Efimov states should appear as real states 
in the spectrum.  Thus, there should be at least one Efimov state 
if $\Lambda^2/m > 6.75 B_2$ and at least two if $\Lambda^2/m > 1406 B_2$.
As an illustration, we consider helium atoms,
which have a large scattering length \cite{Braaten:2002jv}.  
The helium dimer is very shallow: its binding energy $B_2 \approx 1.3$ mK
is smaller than the natural low-energy scale $\Lambda^2/m \approx 400$ mK
by about a factor of 300. Thus we would expect either one or two
Efimov states. There are in fact two helium trimers:
a ground state and an excited state with binding energies 
$B_3' \approx 130$ mK and $B_3 \approx 2$ mK.  Both can be interpreted as 
Efimov states \cite{Braaten:2002jv}. 

The large $D^0 \bar D^{*0}$ scattering length raises the exciting possibility 
of shallow $D^0 D^0 \bar D^{*0}$ molecules 
within 10 MeV of the  $D^0 D^0 \bar D^{*0}$ threshold 
generated by the Efimov effect.
Unfortunately, this possibility can be excluded.
The $D^0 D^0 \bar D^{*0}$ sector involves only
two identical bosons and only two of the three pairs of particles 
have a resonant interaction with a large scattering length.
Furthermore a $D^0 \bar D^{*0}$ pair can fluctuate into a
$\bar D^0 D^{*0}$ pair, and the other $D^0$ has no resonant interaction 
with this component of the wavefunction.
Low-energy interactions in the 3-body sector can again be 
described by an effective potential which in the region 
$m_\pi^{-1} \ll R \ll |a|$  has the scale-invariant form (\ref{Veff}).
The form of the potential can be derived 
from results given in Ref.~\cite{NFJG01}.
If we ignore the 8\% mass difference between the $D^0$ and $\bar D^{*0}$,
the only difference is that the equation for $\lambda_0$ is
\begin{eqnarray}
\sqrt{3} \lambda^{1/2} \cos(\pi \lambda^{1/2}/2) 
= 2 \sin(\pi \lambda^{1/2}/6).
\label{s00}
\end{eqnarray}
Of the factor of 4 difference with (\ref{s0}),
one factor of 2 comes from there being only two 
identical bosons instead of three and the other factor of 2
comes from the 3-body system being a superposition of a
$D^0 D^0 \bar D^{*0}$ molecule and a $D^0 \bar D^0 D^{*0}$ molecule.
The minimum nontrivial solution 
to (\ref{s00}) is $\lambda_0 \approx 0.3533$.
Since this is positive, the Efimov effect does not arise
and there are no shallow 3-body bound states.

\section{A Field Theory Model}
\label{sec:Model}

If we consider only momenta small compared to the natural 
momentum scale $m_\pi$, hadrons such as $D^0$ and $\bar D^{*0}$ 
can be treated as point particles with pointlike interactions
and can therefore
be described by a local nonrelativistic field theory.
In Section~\ref{sec:LEU}, we discussed two fine-tuning mechanisms 
that can generate a large scattering length for 
$D^0 \bar D^{*0}$/$\bar D^0 D^{*0}$. 
Only one of these mechanism is capable of producing a large scattering
length in the $1^{+-}$ channel, but either one is capable of 
producing a large scattering length in the $1^{++}$ channel. 
We will therefore focus on the possibility $J^{PC} = 1^{++}$ 
for the quantum numbers of $X(3870)$.
A model that can describe either of the two fine-tuning mechanisms
is a nonrelativistic field theory with local fields 
$D$, $\bar D$, ${\bf D}$ , $\bar {\bf D}$, and {\boldmath $\chi$} 
for the $D^0$, $\bar D^0$, $D^{*0}$, $\bar D^{*0}$, and $\chi_{c1}(2P)$. 
The hamiltonian density 
is the sum of mass terms, kinetic terms, and interaction terms:
\begin{subequations}
\begin{eqnarray}
{\cal H}_{\rm mass} &=& 
m_{D^0}  \left( D^\dagger D + \bar D^\dagger \bar D \right)
+ m_{D^{*0}} \left( {\bf D}^\dagger \cdot {\bf D} 
	+ \bar {\bf D}^\dagger \cdot \bar {\bf D} \right)
\nonumber
\\
&&+ (m_{D^0}+m_{D^{*0}} + \nu_0) 
	\mbox{\boldmath $\chi$}^\dagger \cdot \mbox{\boldmath $\chi$}
\label{Hm}
\\
{\cal H}_{\rm kin} &=& 
- \mbox{$1\over2$} m_{D^0}^{-1} \left( D^\dagger \nabla^2 D 
	+ \bar D^\dagger \nabla^2 \bar D \right) 
- \mbox{$1\over2$} m_{D^{*0}}^{-1} 
\left( {\bf D}^\dagger \cdot \nabla^2 {\bf D}
	+ \bar {\bf D}^\dagger \cdot \nabla^2 \bar {\bf D} \right)
\nonumber
\\
&&- \mbox{$1\over2$} (m_{D^0}+m_{D^{*0}})^{-1} 
\mbox{\boldmath $\chi$}^\dagger \cdot \nabla^2 \mbox{\boldmath $\chi$},
\\
{\cal H}_{\rm int} &=&
\lambda_0 \left( D \bar {\bf D} + \bar D {\bf D} \right)^\dagger
\cdot \left( D \bar {\bf D} + \bar D {\bf D} \right)
\nonumber
\\
&& + g_0 \left[ \mbox{\boldmath{$\chi$}}^\dagger \! \cdot \! 
	(D \bar {\bf D} + \bar D {\bf D})
+ (D \bar {\bf D} + \bar D {\bf D})^\dagger
\! \cdot \! \mbox{\boldmath $\chi$} \right] ,
\label{Hint}
\end{eqnarray}
\end{subequations}
where $\lambda_0$, $g_0$, and $\nu_0$ are bare parameters
that require renormalization.
A similar field theory has been used to describe the behavior 
of cold atoms near a Feshbach resonance \cite{KMCWH02}.
If we impose an ultraviolet cutoff $\Lambda$ on loop momenta
and drop terms that decrease as inverse powers of $\Lambda$,
the cutoff dependence can be removed by eliminating $\lambda_0$, $g_0$,
and $\nu_0$ in favor of renormalized parameters $\lambda$, $g$,
and $\nu$ defined by
\begin{subequations}
\begin{eqnarray}
\lambda & = & Z_\lambda^{-1} \lambda_0,
\label{lambda}
\\
g & = & Z_\lambda^{-1} g_0,
\label{g}
\\
\nu & = & \nu_0 + [Z_\lambda^{-1}-1] g_0^2/\lambda_0,
\label{nu}
\end{eqnarray}
\label{coeff}
\end{subequations}
where the renormalization constant $Z_\lambda$ is
\begin{equation}
Z_\lambda = 
1 + {2 \over \pi^2} m_{\rm red} \lambda_0 \Lambda 
\label{Z}
\end{equation}
and $m_{\rm red} = m_{D^0} m_{D^{*0}}/(m_{D^0} + m_{D^{*0}})$ is the reduced mass.
Note that the combinations $g_0/\lambda_0 = g/\lambda$ and
$\nu_0 - g_0^2/\lambda_0 = \nu - g^2/\lambda$ 
are renormalization invariants.

The natural scale for the ultraviolet momentum cutoff is 
$\Lambda \sim m_\pi$.  The natural magnitude for
the bare coupling constant $\lambda_0$ can be deduced by 
dimensional analysis:  $|\lambda_0| \sim 1/(m_{\rm red} m_\pi)$.
This can be made evident by writing the renormalization condition  
(\ref{lambda}) in the form
\begin{equation}
{1 \over \lambda} =  {1 \over \lambda_0} 
+ {2 \over \pi^2} m_{\rm red} \Lambda.
\label{lam-inv}
\end{equation}
If the renormalized coupling constant $\lambda$ is fixed
and $\Lambda$ is sufficiently large, $\lambda_0$ must 
scale like $(m_{\rm red} \Lambda)^{-1}$ to compensate 
for the effect of the ultraviolet cutoff.
The natural magnitude for $g_0$ is $\zeta m_\pi^{1/2}/m_{\rm red}$,
where the factor of $m_\pi^{1/2}/m_{\rm red}$ comes from dimensional analysis
and $\zeta$ is a numerical suppression factor associated with the
violation of Zweig's rule by the process $\chi_{c1}(2P) \to D^0 \bar D^{*0}$,
which requires the creation of a light quark-antiquark pair.
The renormalization condition (\ref{g}) implies that the numerical 
suppression factor $\zeta$ is stable under renormalization 
and does not require fine-tuning.
There is no natural magnitude for the bare parameter $\nu_0$:
it is completely adjustable.
In the absence of fine-tuning, the renormalization constant
$Z_\lambda$ in (\ref{Z}) is comparable to 1.
The renormalization conditions (\ref{lambda}), (\ref{g}), 
and (\ref{nu}) then imply that the natural magnitudes 
of the renormalized coupling constants are
$|\lambda| \sim (m_{\rm red} m_\pi)^{-1}$, 
$|g| \sim \zeta m_\pi^{1/2}/m_{\rm red}$,
and $|\nu| \sim {\rm max}(|\nu_0|, \zeta^2 m_\pi^2/m_{\rm red})$.

The 2-body scattering amplitude in this model can be calculated 
analytically.  The scattering length is
\begin{eqnarray}
a &=&  {m_{\rm red} \over \pi} \left( \lambda - {g^2 \over \nu} \right) .
\label{a}
\end{eqnarray}
The natural magnitude for $|a|$ is $1/m_\pi$.
The scattering length can be made unnaturally large either by making 
$\lambda$ sufficiently large, which corresponds to tuning the potential 
between $D^0$ and $\bar D^{*0}$, or by making $\nu$ sufficiently small,
which corresponds to tuning the energy gap
between the $\chi_{c1}(2P)$ and the $D^0 \bar D^{*0}$ threshold.
In either case, low-energy universality implies that 
as $a$ increases,
the binding energy of the molecule approaches (\ref{B2})
and the $D^0 \bar D^{*0}$
or $\bar D^0 D^{*0}$ wavefunction approaches (\ref{psi}).

The first mechanism for generating a large scattering length
is to make $\lambda$ unnaturally large: $|\lambda| \gg |\lambda_0|$.
This can be accomplished by tuning $\lambda_0$ 
towards the critical value $-\pi^2/(2 m_{\rm red} \Lambda)$,
so that there is a near cancellation between the two terms 
on the right side of (\ref{lam-inv}).
This fine-tuning makes the renormalization constant  
$Z_\lambda$ much less than 1.
The renormalization condition (\ref{g}) implies that this fine-tuning
also increases the strength of the effective coupling constant
between {\boldmath{$\chi$}} and $D \bar {\bf D}$: $|g| \gg |g_0|$.
This is also evident from the fact that $g/\lambda = g_0/\lambda_0$
is a renormalization invariant.
There is a limit to how large the scattering length can be made using
this mechanism.  When $Z_\lambda$ becomes smaller than $g_0^2/|\lambda_0 \nu_0|$,
the $g_0^2/\lambda_0$ term in (\ref{nu}) begins to dominate over the
$\nu_0$ term.  In this case, both terms in the scattering length (\ref{a})
become large and they tend to cancel each other.
Thus, with this mechanism, the maximum magnitude of the scattering length 
is of order $(\lambda_0/g_0)^2 m_{\rm red} |\nu_0|$ which is of order
$\zeta^{-2}m_{\rm red} |\nu_0|/m_\pi^3$.

The second mechanism for generating a large scattering length
is to make $\nu$ sufficiently small.
This can be accomplished by tuning $\nu_0$ towards the critical value
$-[Z_\lambda^{-1}-1] g_0^2/\lambda_0$
for which there is a near cancellation between the two terms 
on the right side of (\ref{nu}).
The scattering length can be made arbitrarily large using this mechanism.

The calculation of the binding energy $B_2$ of $X$ 
and of the probability $Z_\chi$ 
for the $\chi_{c1}(2P)$ component of the wavefunction
can both be reduced to the solution of a cubic polynomial.
The binding momentum $\kappa$ defined by $B_2 = \kappa^2/(2 m_{\rm red})$
satisfies the cubic equation
\begin{equation}
\kappa^2 + 2 m_{\rm red}  \nu = {m_{\rm red} \over \pi} \lambda \kappa 
\left[ \kappa^2 + 2 m_{\rm red} \left( \nu - g^2/\lambda \right) \right].
\end{equation}
In either of the two limits $\lambda \to \infty$ 
or $\nu \to 0$, one of the three roots of this polynomial 
has the limiting behavior $\kappa \to 1/a$.
If $a>0$, the probability $Z_\chi$ for the $\chi_{c1}(2P)$ component 
of the wavefunction is
\begin{equation}
Z_\chi = {1 \over 2 \pi}
\left ( { g^2 /\lambda^2 \over [\kappa^2 + 2 m_{\rm red} (\nu- g^2 /\lambda)]^2}
	+ {1 \over 4 \pi \kappa} \right)^{-1}
{\kappa - \pi/(m_{\rm red} \lambda) \over \kappa^2 + 2 m_{\rm red} (\nu - g^2/\lambda)}.
\end{equation}
After expressing the observables $B_2$ and $Z_\chi$ as functions of 
$a$ and the renormalization invariants $g/\lambda$
and $\nu - g^2/\lambda$, they can be expanded in powers of $1/a$:
\begin{subequations}
\begin{eqnarray}
B_2 & \approx & {1 \over 2 m_{\rm red} a^2} 
\left( 1 -  {\pi (g/\lambda)^2 \over m_{\rm red}^2 (\nu - g^2/\lambda)^2 a} + \ldots \right),
\\
Z_\chi & \approx &  
{\pi (g/\lambda)^2 \over m_{\rm red}^2 (\nu - g^2/\lambda)^2 a} + \ldots.
\end{eqnarray}
\end{subequations}
For any fine-tuning that produces a large scattering length,
the bare coupling constants approach definite limiting values
and therefore the renormalization invariants 
$g/\lambda$ and $\nu - g^2/\lambda$ approach definite limiting values.
Thus the probability $Z_\chi$ decreases like $1/a$.
This illustrates our assertion that with the probability 
for states other than 
$D^{*0} \bar D^{*0}$ or $\bar D^{*0} D^{*0}$ scales as $1/a$.

We proceed to discuss how the decays of $X(3870)$
could be described within this effective field theory.
In the decay $X \to D^0 \bar D^0 \pi^0$, which is dominated by the 
decay of a constituent $D^{*0}$ or $\bar D^{*0}$, 
the typical momentum of the final $D^0$ or $\bar D^0$ is 40 MeV,
which is much smaller than the natural momentum scale $m_\pi$.
Thus this decay can be described within the effective theory 
by introducing a $\pi^0$ field into the lagrangian with an 
interaction term that allows the decay $\bar D^{*0}\to D^0 \pi^0$.
In the decay $X \to D^0 \bar D^0 \gamma$, which is also dominated by 
the decay of a constituent $D^{*0}$ or $\bar D^{*0}$, 
the typical momentum of the recoiling $D^0$ or $\bar D^0$ is 140 MeV,
which is comparable to the natural momentum scale $m_\pi$.
Thus this decay need not be described accurately within an 
effective theory in which hadrons are treated as point particles 
with pointlike interactions.

The radiative and hadronic transitions and the annihilation decays
of $X(3870)$ produce particles with momenta larger than 
the $m_\pi$.  They therefore cannot be described explicitly
within an effective theory in which hadrons are treated 
as point particles with pointlike interactions.
The inclusive effects of these decays can however be taken into account 
implicitly through local terms in the hamiltonian density.   
The inclusive effects of transitions of
$D^0 \bar D^{*0}$ or $\bar D^0 D^{*0}$ to charmonium states
and of their annihilation into light hadrons
can be taken into account through an 
imaginary part of the bare coupling constant $\lambda_0$.
The inclusive effects of transitions of $\chi_{c1}(2P)$
to other charmonium states and of its decays into light hadrons
can be taken into account through an 
imaginary part of the bare parameter $\nu_0$:
${\rm Im}\nu_0 = - {1\over2}\Gamma_{\chi_{c1}(2P)}$.
The imaginary part of $g_0$ can take into account
interference effects associated  with transitions of $D^0 \bar D^{*0}$
and $\chi_{c1}(2P)$ to the same final states.
If the parameters $\lambda_0$, $g_0$, and $\nu_0$ have small 
imaginary parts, the scattering length (\ref{a}) is complex-valued 
with a small imaginary part.  
If a fine-tuning makes the real part of $a$ large,
the binding energy of $X$ is given by the real part of the expression 
(\ref{B2}).  Its imaginary part should be interpreted as 
${1\over2}\Delta \Gamma_X$, where $\Delta \Gamma_X$
is the contribution to the width from the decays
whose effects are taken into account through Im$\lambda_0$, Im$g_0$, 
and Im$\nu_0$. 
At first order in the imaginary parts of $\lambda_0$, $g_0$, and $\nu_0$,
the contribution to the width is
\begin{equation}
\Delta \Gamma_X = {2 \over \pi a^3}
\left[ (1 -2 a \Lambda /\pi)^2 (-{\rm Im}\lambda_0) 	
	+ 2 {g \over \nu} (1 -2 a \Lambda /\pi){\rm Im}g_0
	+ {g^2 \over \nu^2} (-{\rm Im}\nu_0) \right] .
\label{dGamX}
\end{equation}
If we express $g/\nu$ in terms of $a$ and the renormalization
invariants, we see that it increases linearly with $a$:
$g/\nu = a(g/\lambda)/(\nu - g^2/\lambda)$.  Thus all three terms 
in (\ref{dGamX}) scale as $1/a$ in the limit $a \to \infty$.
This scaling behavior is in agreement with that deduced 
in Section~\ref{sec:Decays}.


\section{Summary}
\label{sec:Summary}

We have explored the implications of low-energy universality for the
identification of $X(3870)$ as 
an S-wave  $D^0 \bar D^{*0}$/$\bar D^0 D^{*0}$ molecule.
Its shallow binding energy requires some fine-tuning
mechanism to generate a large $D^0 \bar D^{*0}$ scattering length $a$.  
Two possible mechanisms are an accidental fine-tuning of parameters 
associated with the $D^0 \bar D^{*0}$ sector and an accidental fine-tuning 
of a P-wave charmonium state to the $D^0 \bar D^{*0}$ threshold.  
A field theory model that illustrates both of these mechanisms was presented.  
With either mechanism, the probabilities for components of the wavefunction 
other than $D^0 \bar D^{*0}$ or $\bar D^0 D^{*0}$ 
are suppressed by a factor of $1/a$.  The decay rates into modes 
other than those associated with decay of a constituent 
$\bar D^{*0}$ or $D^{*0}$ are also suppressed by a factor of $1/a$.
The assumption that the large scattering length arises from the 
fine-tuning of $\chi_{c1}(2P)$ or $h_c(2P)$ to the $D^0 \bar D^{*0}$ threshold 
is particularly predictive.  The decay rates 
for radiative transitions and for hadronic transitions 
via emission of light hadrons with total isospin 0
should differ from the corresponding decay rates 
of $\chi_{c1}(2P)$ or $h_c(2P)$ in the absence of the fine-tuning by a 
common suppression factor.  Low-energy universality 
also has nontrivial predictions for 3-body systems, such as 
$D^0 D^0 \bar D^{*0}$, although the spectacular possibility
of Efimov states can be excluded. 
In conclusion, if the $X(3870)$ is indeed an S-wave 
$D^0 \bar D^{*0}$/$\bar D^0 D^{*0}$ molecule, 
it will provide a beautiful 
example of the remarkable phenomenon of low-energy universality.

This research was supported in part by the Department of Energy under
grant DE-FG02-91-ER4069.


\end{document}